\documentclass[twoside,12pt]{article}
\usepackage{epsfig}

\newcommand{\be}{\begin{equation}}
\newcommand{\ee}{\end{equation}}
\newcommand{\bea}{\begin{eqnarray}}
\newcommand{\eea}{\end{eqnarray}}

\begin{document}

\title{ \vspace{1cm} Microscopic Cluster Model for Exotic Nuclei}
\author{M.\ Tomaselli,$^{1,2}$ T. K\"uhl,$^2$ D.\ Ursescu,$^2$ \\
$^1$TUD, Technische Universit\"at Darmstadt, D-64289 Darmstadt, Germany\\
$^2$GSI, Gesellschaft f\"ur Schwerionen Forschung, D-64291 Darmstadt, Germany}

\maketitle
\begin{abstract} 
For a better understanding of the dynamics of exotic nuclei it is 
of crucial importance to develop a practical microscopic theory easy
to be applied to a wide range of masses.
Theoretically the basic task consists in formulating an easy solvable theory
able to reproduce structures and transitions of known nuclei which
should be then used to calculate the sparely known properties of proton-
or neutron-rich nuclei. In this paper we start by calculating energies
and distributions of $A \leq 4$ nuclei withing a unitary correlation model
 restricted to include only two-body correlations.
The structure of complex nuclei is then calculated extending the
model to include correlation effects of higher order.     
\end{abstract}
\section{Introduction}
The experimental efforts presently performed at the new generation
of radioactive ion beam facilities in particular the Rare Isopote Accelerator
(RIA) in USA, the international accelerator facility for research with
anti-proton ions at GSI in Germany, and the Radioactive Ion Beam Factory 
(RIBF) in Japan, will boost the nuclear structure studies of exotic nuclei.

One of the central challenges of theoretical nuclear physics is the attempt to
describe these unknown properties of the exotic systems in terms of a realistic
nucleon-nucleon (NN) interaction.
In order to calculate matrix elements with this singular interaction
we have to define effective correlated Hamiltonians. 

Correlation effects in nuclei have been first introduced in nuclei by Villars~\cite{vil63},
who proposed the unitary-model operator (UMO) to construct effective operators.
The method was implemented by Shakin~\cite{sha66} for the calculation
of the G-matrix from hard-core interactions.
Non perturbative approximations of the UMO have been recently applied to odd nuclei 
in Ref.~[3] and to even nuclei in Ref.~[4]. 
The basic formulas of the Dynamic Correlation Model and of the
 Boson Dynamic Correlation Model (BDCM) presented in the above
quoted papers have been obtained by separating the n-body correlation 
operator in short- and long-range components.
 The short-range component is considered up to the two body correlation
while for the long range component 
the three and four body correlation operators have been studied.
The extension of the correlation operator to high order diagrams is especially important
in the description of exotic nuclei (open shell).  
In the short range approximation 
the model space of two interacting particles is separated in two subspaces:
one which includes the shell model states and the other (high momentum) which
 is used to compute the G-matrix of the model.
 The long range component of the correlation operator 
 has the effect of generating a new correlated model space (effective space)
which departs from the originally adopted one (shell model).
The amplitudes of the model wave functions are calculated in terms of
non linear equation of motions (EoM).
 The derived systems of commutator equations,
which characterize the EoM, are finally linearized.
Within these generalized linearization approximations (GLA)
we include in the calculation presented in the paper up to the ((n+1)p1h) effective diagrams.
The linearized terms provide, as explained later in the text, the 
additional matrix elements that convert the perturbative UMO expansion 
in an eigenvalue equation. The n-body matrix elements needed to diagonalize the 
 resulting eigenvalue equations are calculated exactly 
via the Cluster Factorization Theory (CFT). \\ 

Within the present treatment of the correlation operator one generates 
in the n-body theory not only the
ladder diagrams of Ref.~\cite{bru01} but also the folded diagrams of Kuo see Ref.~\cite{kuo01}.

\section{The Two-body Effective Interaction}
In order to describe the structures and the distributions of nuclei
we start from the following Hamiltonian:
\begin{equation}\label{equ.1}
H= \sum_{\alpha\beta}
 \langle\alpha|t|\beta\rangle \, a^{\dagger}_{\alpha}a_{\beta}
   \:+\: \sum_{\alpha\beta\gamma\delta}
   \langle\Phi_{\alpha\beta}| v_{12}|\Phi_{\gamma\delta}\rangle \,
   a^{\dagger}_{\alpha}a^{\dagger}_{\beta} a_{\delta}a_{\gamma}  
\end{equation}
where $v_{12}$ is the singular nucleon-nucleon two body potential.
Since the matrix elements $|\alpha\beta\rangle$ are uncorrelated the matrix elements of 
$v_{12}$ are infinite. This problem can be avoided by taking matrix elements
of the Hamiltonian between correlated states.
In this paper the effect of correlation is introduced via the $e^{iS}$ method.
In dealing with very short range correlations only the $S_2$ part of the correlation
 operator needs to be considered.  

Following Ref.~[2] we therefore calculate an ``effective Hamiltonian'' by using only the $S_2$ 
correlation operator obtaining:
\begin{equation}\label{eq.1}
\begin{array}{l}
H_{eff}=e^{-iS_2}He^{iS_2}=\sum_{\alpha\beta}\langle\alpha|t|\beta\rangle a^{\dagger}_{\alpha}
a_{\beta}+
\sum_{\alpha\beta\gamma\delta}\langle\Psi_{\alpha\beta}|v^l_{12}|\Psi_{\gamma\delta}\rangle 
a^{\dagger}_{\alpha}a^{\dagger}_{\beta} a_{\delta}a_{\gamma}\\
=\sum_{\alpha\beta}\langle\alpha|t|\beta\rangle a^{\dagger}_{\alpha}a_{\beta}+
\sum_{\alpha\beta\gamma\delta}\langle\Psi_{\alpha\beta}|v|\Psi_{\gamma\delta}\rangle a^{\dagger}_{\alpha}a^{\dagger}_{\beta} a_{\delta}a_{\gamma}
\end{array}
\end{equation}
where $v_{12}^l$ refers to the long-range part of the nucleon-nucleon force
diagonal in the relative orbital angular momentum,
after the separation~\cite{mos60}:
\begin{equation}
\label{eq.2}
v_{12}=v_{12}^s+v_{12}^l 
\end{equation}
The separation is made in such a way that the short range part produces no
energy shift in the pair state~\cite{mos60}.
In doing shell model calculation with the Hamiltonian Eq.~(\ref{eq.1}),
we remark: a) only the long tail potential plays an essential role in the calculations
of the nuclear structure i.e.: the separation method and the new
proposed $v_{low-k}$~\cite{jia01} method show a strong analogy and
b) the $v_T^{od}$ must be included as an additional re-normalization of the
effective Hamiltonian Eq.~(\ref{eq.1}).
 
In Eq.~(\ref{eq.1}) the $\Psi_{\alpha\beta}$ is the two particle correlated wave function:
\begin{equation}\label{eq.1a}
\Psi_{\alpha\beta}=e^{iS_2}\Phi_{\alpha\beta}
\end{equation}

In order to evaluate the effect of the tensor force on the $\Psi_{\alpha\beta}$
 we calculate:
\begin{equation}
\label{eq.3}
w(r)=V_T^{od}\frac{Q}{\Delta E}u(r)=
V_T^{od}\frac{Q}{\Delta E}|(\tilde{nl}S),J':NL,J\rangle
\end{equation}
where Q is a momentum dependent projection operator, $\Delta E(k_1,k_2)$ 
 the energy denominator and $\tilde{nl}$ the correlated two particle
state in the relative coordinates.
In Eq.~(\ref{eq.3}) u(r) is generated as in Ref.~\cite{sha66} by a separation distance
 calculation for the central part of the force in the $^3S_1$ state.
The wave function obtained in this way (full line) heals to the harmonic-oscillator
 wave function (dashed line) as shown in Fig.~1. 
The result obtained for Eq.~(\ref{eq.3}) calculated with the tensor force 
of the Yale potential~\cite{las62}
is given also in Fig.~1 left where we plot  
for the harmonic oscillator size parameter b=1.41~fm:
\begin{equation}
\label{eq.4}
\Psi(\vec{r})=[u(r)Y^1_0(\Omega_{\vec{r}})+w(r)Y^1_2 (\Omega_{\vec{r}})] 
\end{equation}
Being the admixture of the two components, circa $4\%$, 
the wave function Eq.~(\ref{eq.4}) can
be associated to the deuteron wave function.

Let us use the the Hamiltonian Eq.~(\ref{eq.1}) to calculate the structure of
the A=3 nuclei.
Here we propose to calculate the ground state of $^3$H, $^3$He, and $^4$He 
within the EoM
 method which derive the eigenvalue equations by working with the $e^{iS_2}$ 
operator on the wave functions of the A=3,4 nuclei.

\noindent
\begin{minipage}[hb]{0.48\linewidth}
\epsfig{file=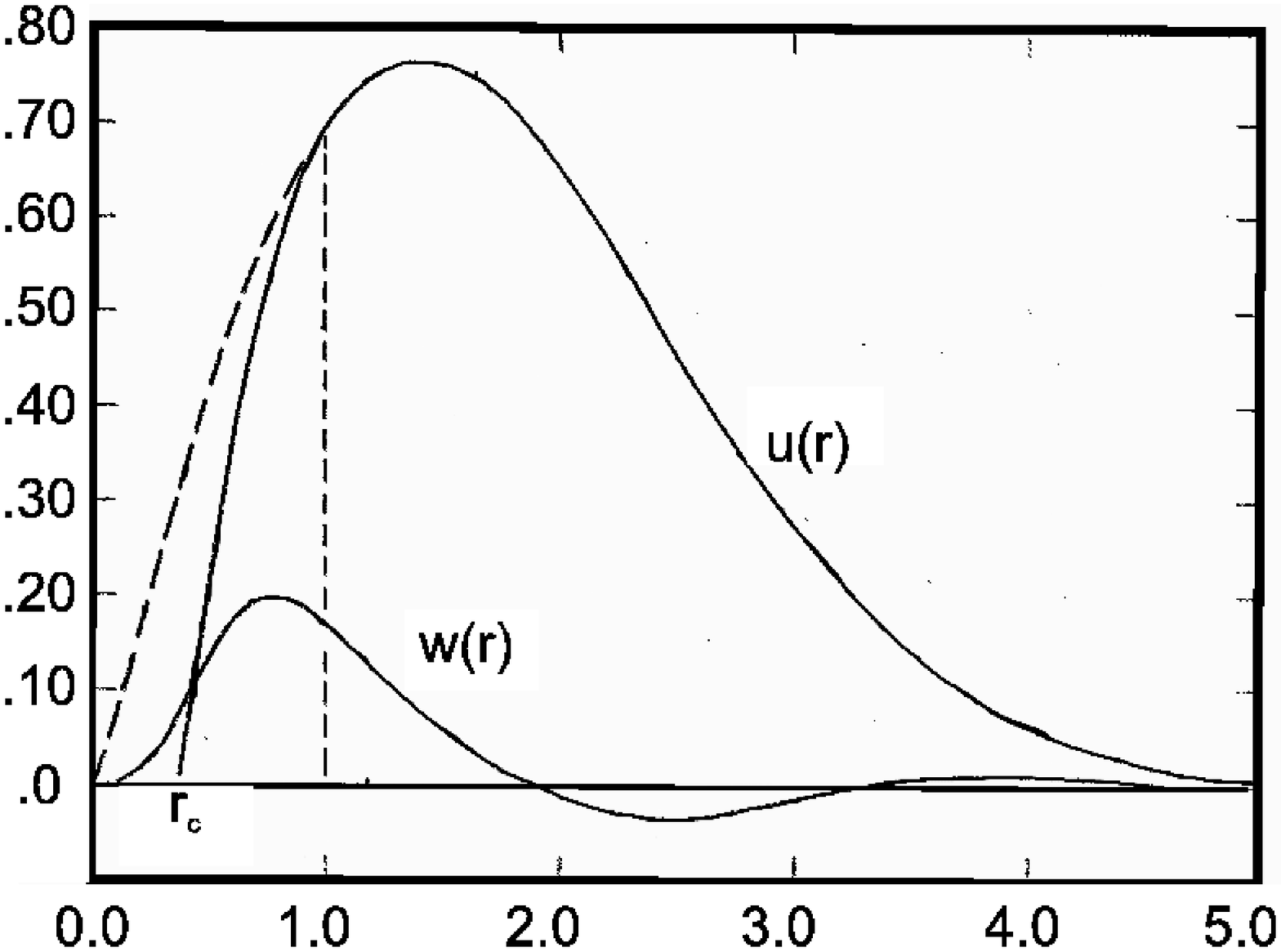,height=6.cm,width=6.5cm,clip=}
\end{minipage}\hfill
\begin{minipage}[hb]{0.48\linewidth}
\epsfig{file=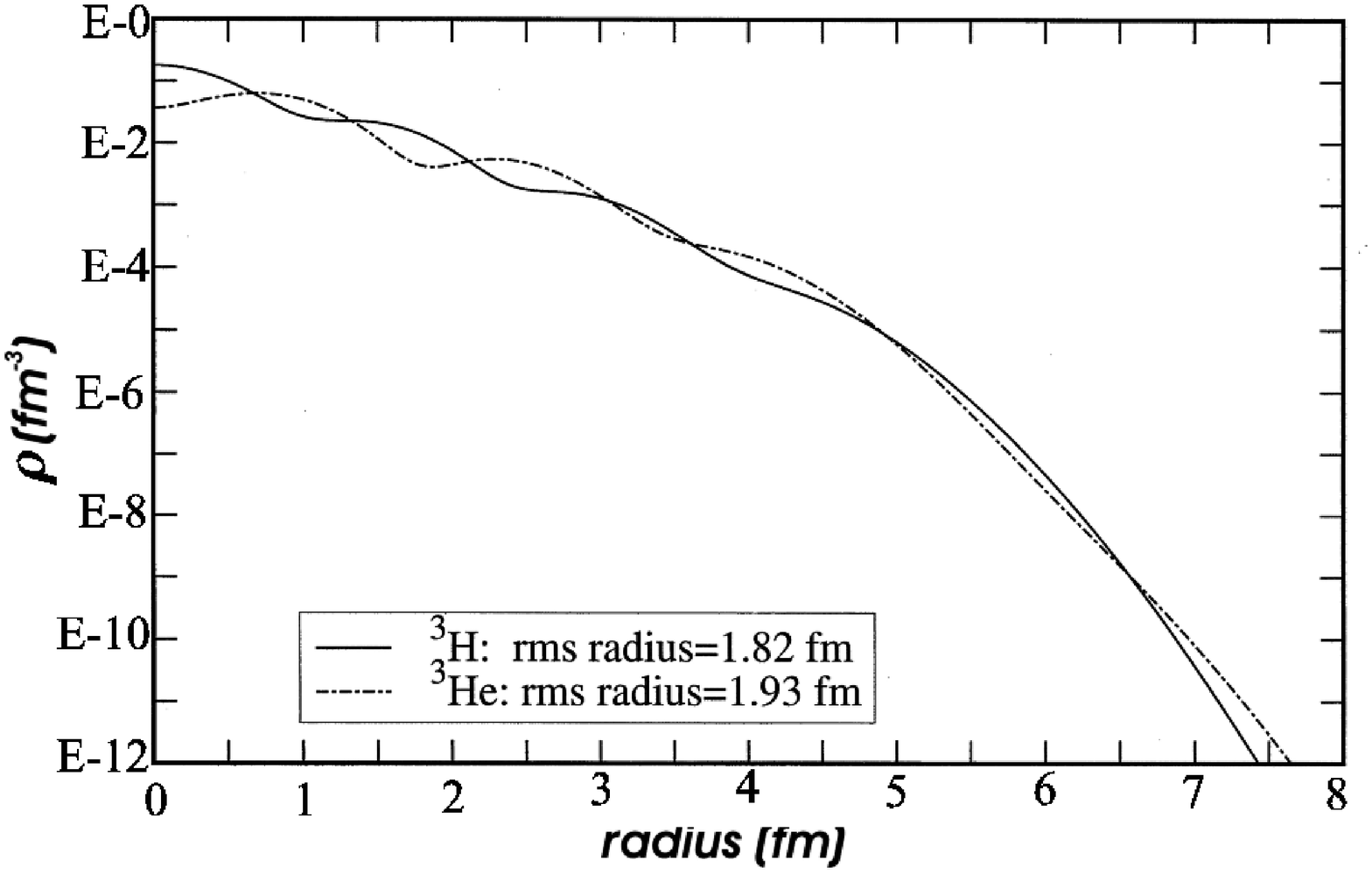,height=6.cm,width=6.5cm}
\end{minipage}
\begin{figure}[htp]
\caption{Left: The u(r) and w(r) wave functions of the deuteron, with
quantum numbers $^3S_1$ and $^3D_1$, plotted as function of r;
Right: Distributions of $^3$H and $^3$He.}
\end{figure}\newline
\section{The Few-body Problem.} 
We write the three particle states in second quantization by discarding for
simplicity the isospin quantum numbers: 
\begin{equation}\label{e1}
\Phi_{3p}\longrightarrow A^{\dagger}_1(\alpha_1J_1J)|0\rangle=
[a^{\dagger}_{j_1}(a^{\dagger}_{j_2}a^{\dagger}_{j_3})^{J_1}]^J_M|0\rangle,
\end{equation}
where the operators $a^{\dagger}_{j_1}a^{\dagger}_{j_2}a^{\dagger}_{j_3}$ create three 
particles in the open shells and we analyze the structure of
the particle dynamics, generated by the correlation operator, via the following commutator:
\begin{equation}\label{e2}
\begin{array}{l}
\displaystyle{ 
[H,A^{\dagger}_1(\alpha_1J_1J)]|0\rangle=
[(\sum_{\alpha}\epsilon_{\alpha}a^{\dagger}_{\alpha}a_{\alpha}+\frac{1}{2}
\sum_{\alpha \beta \gamma \delta} \langle\alpha\beta|v(r)|\gamma\delta\rangle a^{\dagger}_{\alpha} 
a^{\dagger}_{\beta}a_{\delta}a_{\gamma}),(a^{\dagger}_{j_1}
(a^{\dagger}_{j_2}a^{\dagger}_{j_3})^{J_1})^J]|0\rangle}.
\end{array}
\end{equation}
By taking the expectation value of the commutator Eq.~(\ref{e2}) between
the vacuum and the three particle states we obtain the eigenvalue 
equation.
In order to solve Eq.~(\ref{e2}) we introduce the CFT~\cite{tom06}
which factorizes the three particle states in combination of pairs.
For the two particles we distinguish between two spaces:
1) effective valence space which is used to diagonalize the EoM and
2) complementary high excited single particle space which is used to compute
the G matrix.
Within this method we can use either the Shakin-Yale matrix elements 
of Ref.~\cite{sha67} with b=1.50 fm or the $V_{low-k}$ matrix elements~\cite{jia01}.
From the diagonalization of the eigenvalue equation of the three particles, 
we obtain an energy difference
$\Delta E(^3H-^3He)$=0.78 MeV and the charge distributions and radii given in Fig.~1 Right.
 By extending the commutator of Eq.~(\ref{e2}) to a four particle state 
we obtain for the ground state of $^4$He the binding energy of E=28.39 Mev.
From the the ground state wave functions calculated with the two-body potential of 
Ref.~\cite{sha67} (b=1.50 fm) and  with the $V_{low-k}$ matrix 
elements~\cite{jia01} evaluated with the Bonn potential~\cite{mal89} 
we obtain the two distributions given in Fig.~2) left. 
The distributions and the radii of 1.709 fm and 1.71 fm look similar.
 
\noindent
\begin{minipage}[hb]{0.48\linewidth}
\epsfig{file=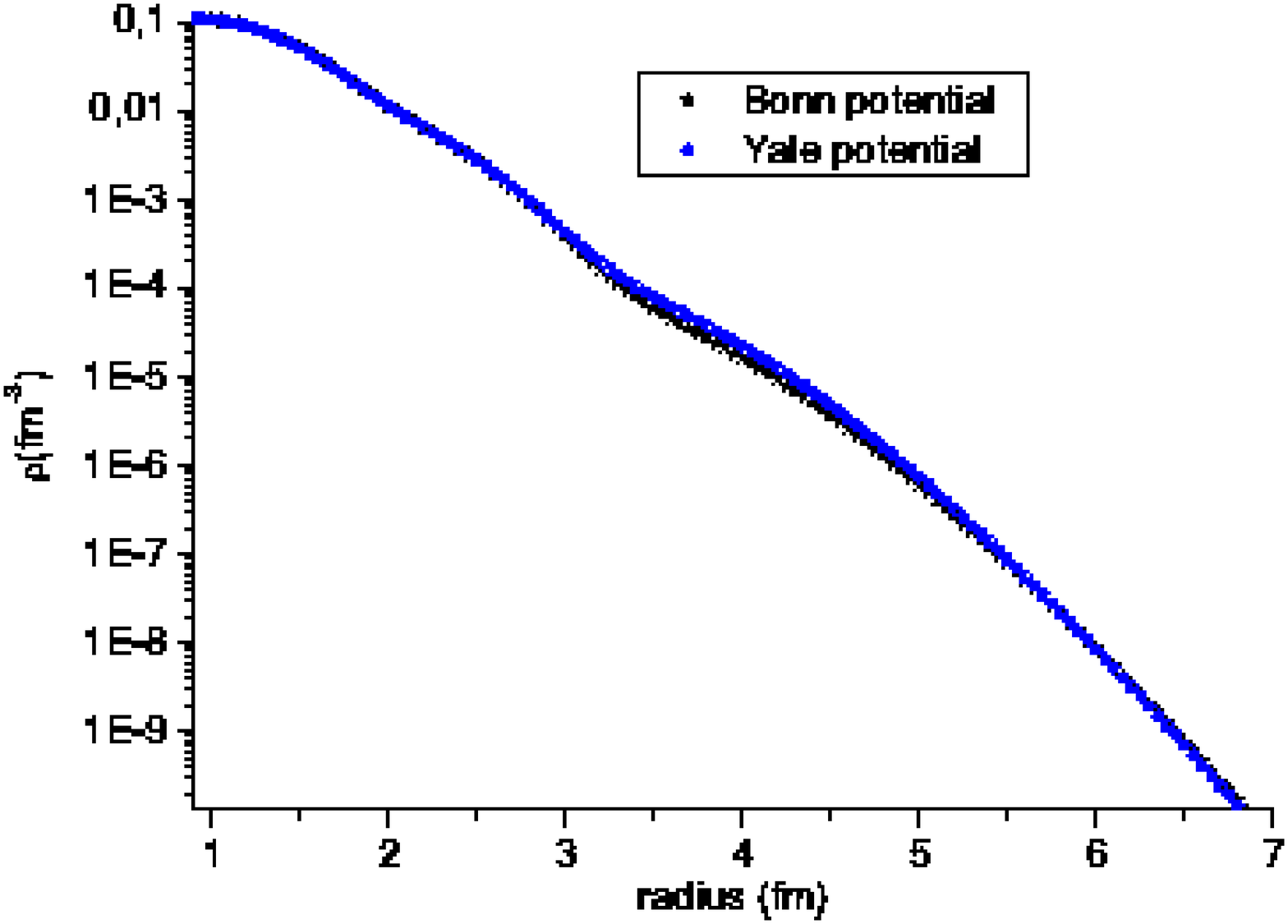,angle=0.,height=6.cm,width=6.5cm,clip=}
\end{minipage}\hfill
\begin{minipage}[hb]{0.48\linewidth}
\epsfig{file=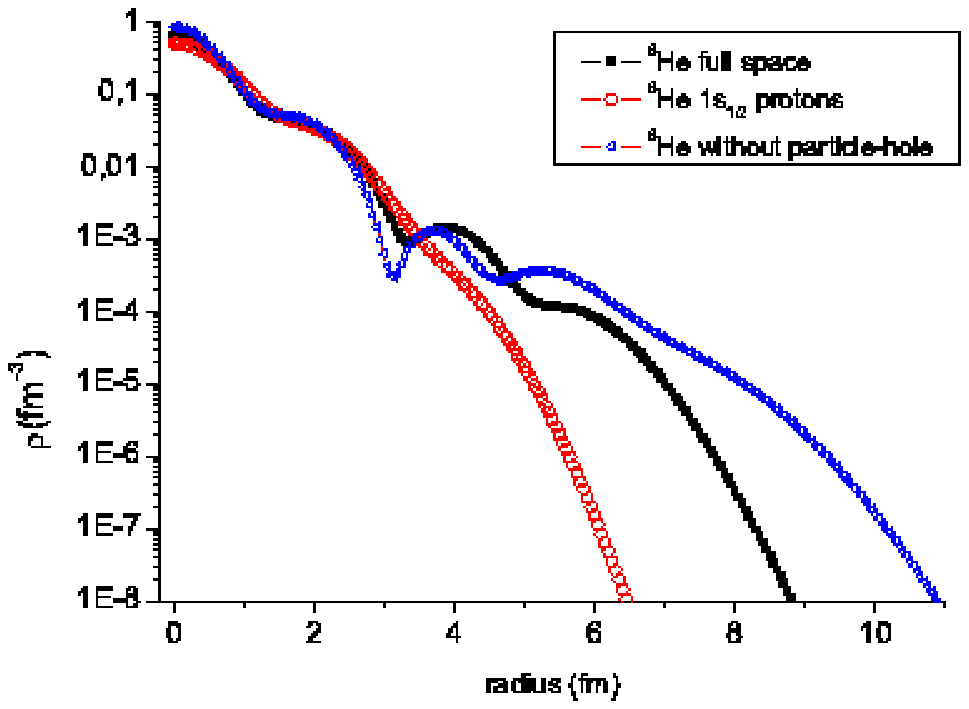,angle=0.,height=6.cm,width=6.5cm,clip=}
\end{minipage}\vspace{-1ex}
\begin{figure}[htp]
\caption{Left: Distributions of $^4$He calculated with the Shakin-Yale and the 
$V_{low-k}$ potentials;
Right: Charge distributions of $^6$He calculated in different approximations.}
\end{figure}\newline
\section{The n-Body Problem} 
 In dealing with complex nuclei however the ($S_i,~i=3\cdots n$)
correlations should also be considered.
The evaluation of these diagrams is, due to the 
exponentially increasing number of terms, difficult in a perturbation theory.
We note however that one way to overcome this problem is to work with 
$e^{i(S_1+S_2+S_3+\cdots+S_i)}$ operator on the Slater's determinant
 by keeping the n-body Hamiltonian uncorrelated.
Via the long tail of the nuclear potential the Slater determinant of the 
``n'' particle systems are interacting with the excited Slater's determinants
formed by the (``n'' particles+(mp-mh) mixed-mode excitations). 
The amplitudes of the different determinants are calculated via
the EoM method.

After having performed the diagonalization of the n-body Hamilton's operator we
can calculate the form of the effective Hamiltonian which, by now, includes 
the complete set of the commutator equations.
For odd particle systems we use the well established formalism of Ref.~[3].
\subsection{Two ``dressed'' Particles}
In the following we present the theory for even ``dressed'' particles.
In order to define the eigenvalue equations for the nuclear modes we start
by defining the two particle operator as following:
\begin{equation}\label{e21}
\Phi_{2p}\longrightarrow A^{\dagger}_1(\alpha_1J)|0\rangle=[a^{\dagger}_{j_1}a^{\dagger}_{j_2}]^J_M|0\rangle,
\end{equation}
 We calculate then the commutator equations:
\begin{equation}
\label{e34}
[H,A^{\dagger}_1(\alpha_1J)]|0\rangle=\sum_{\beta_{1}} \Omega(2p|2p') A^{\dagger}_1(\beta_1J)|0\rangle+ 
\sum_{\beta_2J'_1J'_2} \Omega(2p|3p1h) A^{\dagger}_2(\beta_2J'_1J'_2J)|0\rangle.
\end{equation}
and 
\begin{equation}
\label{e8}
\begin{array}{l}
\displaystyle{
 [H,A^{\dagger}_2(\alpha_2J_1J_2J)]|0\rangle}  \\
\displaystyle{
=\sum_{\beta_{2}J'_1J'_2} \Omega(3p1h|3p'1h') A^{\dagger}_2(\beta_2J'_1J'_2J)|0\rangle+
\sum_{\beta_{3}J'_1J'_2J'_3} \Omega (3p1h|4p2h) A^{\dagger}_3(\beta_3J'_1J'_2J'_3J)|0\rangle},
\end{array}
\end{equation}
which define the dynamic evolution of the valence modes.
In Eqs.~(\ref{e34},\ref{e8}) the $A^{\dagger}_2(\beta_2J'_1J'_2J)$ operators are defined below:
\begin{equation}\label{e5}
\begin{array}{l}
\Phi_{3p1h}\longrightarrow A^{\dagger}_2(\beta_2J'_1J'_2J)|0\rangle
=((a_{j'_1}^{\dagger}a_{j'_2}^{\dagger})^{J'_1}(a_{j'_3}^{\dagger}a_{j'_4})^{J'_2})^J|0\rangle.
\end{array}
\end{equation}
The additional commutator equations which involves the $A^{\dagger}_3(\beta_3J'_1J'_2J'_3J)$
and the higher order operators are here not given. 
The obtained commutator chain is suitable to be solved perturbatively by inserting 
the $n$-th commutator in the $(n-1)$-th commutator, $(n-1)$-th commutator 
in the $(n-2)$-th commutator 
, \ldots the second commutator in the first.
Within this perturbative approach one defines effective Hamiltonians 
of the model which, due to the increasing degree of complexity, are not easily solvable.
Much simpler solutions to the commutator equations may, however, be obtained in
the BDCM model.
We start by remarking that in the study of low lying excitations of the
n-body systems the higher order components of the wave functions, which involve 
 n valence - and (2p-2h) core-excitations are poorly admixed in the
model space and can be linearized. 
 This linearization consists by 
applying the Wick's theorem to the $A^{\dagger}_3(\beta_3J'_1J'_2J'_3J)$ terms 
and by neglecting the normal order diagrams.
The linearization generates the additional terms that one
needs to convert the commutator chain in the corresponding eigenvalue equation,
as can be obtained by taking the expectation value of the linearized Eqs.~(\ref{e34}),
and~(\ref{e8}) between the vacuum and the model states.
\subsection{Three ``dressed'' Particles}
For three ``dressed'' particles we extend the commutator equation of Eq.~(\ref{e1})
obtaining the following non-linear commutator equations:
\begin{equation}
\label{e35}
[H,A^{\dagger}_1(\alpha_1J_1J)]|0\rangle=\sum_{\beta_{1}} \Omega(3p|3p') 
A^{\dagger}_1(\beta_1J'_1)|0\rangle+ 
\sum_{\beta_2J'_1J'_2} \Omega(3p|4p1h) A^{\dagger}_2(\beta_2J'_1J'_2J)|0\rangle.
\end{equation}
and 
\begin{equation}
\label{e35a}
\begin{array}{l}
\displaystyle{
 [H,A^{\dagger}_2(\alpha_2J_1J_2J)]|0\rangle}  \\
\displaystyle{
=\sum_{\beta_{2}J'_1J'_2} \Omega(3p|3p') A^{\dagger}_2(\beta_2J'_1J'_2J)|0\rangle+
\sum_{\beta_{3}J'_1J'_2J'_3} \Omega (3p|4p2h) A^{\dagger}_3(\beta_3J'_1J'_2J'_3J)|0\rangle},
\end{array}
\end{equation}
In Eqs.~(\ref{e35},\ref{e35a}) 
the $A^{\dagger}_2(\beta_2J'_1J'_2J)$ operators are defined below:
\begin{equation}\label{e35b}
\begin{array}{l}
\Phi_{3p1h}\longrightarrow A^{\dagger}_2(\beta_2J'_1J'_2J)|0\rangle
=((a_{j'_1}^{\dagger}(a_{j'_2}^{\dagger}a_{j'_3}^{\dagger})^{J'_1})^{J'_i}
 (a_{j'_5}^{\dagger}a_{j'_4})^{J'_2})^J|0\rangle.
\end{array}
\end{equation}
The additional commutator equations are here not given. 
 The $(A^{\dagger}_3(\beta_3J'_1J'_2J'_3J)$ (4p1h) terms in Eq.~(\ref{e35b}) are 
finally linearized. By neglecting the normal order terms we obtaing
the terms which  convert the commutator chain in
coupled non-linear equations for
three particles interanting with the (4p1h) states.  
The eigenvalue equation for the three dressed particles
is then obtained by taking the expectation value of the linearized Eqs.~(\ref{e35}),
and~(\ref{e35a}) between the vacuum and the excited states.

The generalization of the commutator equations to ``n'' valence particles 
can be simply derived in the analogous way and is not given. 
Within these approximations the model 
commutator equations are suitable to be restricted to a finite space. 
The linearized system of the commutator equations is then
solved exactly in terms of the CFT~\cite{tom06} which calculates the n-body
matrix elements of one- and two-body operators.
\section{Results}
In order to perform structure calculations, we have to define 
the CMWFs base, the ``single-particle energies'' and to choose the
 nuclear two-body interactions.
The CMWFs are defined as shown in Appendix A by including mixed valence modes
and core-excited states.
The base is then orthonormalized and, since the single particle wave functions
 are harmonic oscillators, the center-of-mass (CM) is removed. 
 
One can generate the CM spurious states according to Ref.~\cite{bar61}
and evaluate the overlap between these states and the nuclear
eigenstates of the model (see Appendix B).
Model components having with the corresponding CM components an overlap greater than 10\% 
are treated as spurious states and discarded.
The single-particle energies of these levels are taken from the known experimental
level spectra of the neighboring nuclei~\cite{bro01}.  
For the particle-particle interaction, we use the G-matrix obtained from
 Yale potential~\cite{sha67}.
These matrix elements are evaluated by applying the
$e^S$ correlation operator, truncated at the second order
term of the expansion, to the harmonic oscillator
base with size parameter b=1.76 fm.   
As elucidate in Refs.~[3] and [4] the effective two-body
 potential used by the DCM and the BDCM models
is separated in low and high momentum components.
 Therefore, the effective model matrix elements calculated within 
 the present separation method
and those calculated by Kuo~\cite{jia01} in the $v_{low-k}$ approximation
 are pretty similar.
The adopted separation method and the $v_{low-k}$ generate two-body
matrix elements which are almost
independent from the radial shape of the different potentials generally
used in structure calculations. \\
The particle-hole matrix elements could be calculated from the particle-particle 
matrix elements via a re-coupling transformation. 
We prefer to use the phenomenological potential of Ref.~\cite{mil75}. 
The same size parameter as for the particle-particle
 matrix elements has been used.
In this contribution we present application of the $S_n$ correlated model to the
charge distribution of $^6$He and to the electromagnetic transitions
of neutron rich Carbon and Oxygen isotopes.
In Fig.~2) Right three distributions are given for $^6$He: 
1) the correlated charge distribution  
calculated with the full $S_3$ operator, 
2) the correlated charge distribution  
calculated with the partial $S_3$ operator obtained by neglecting the folded diagrams, 
3) the charge distribution calculated 
for two correlated protons in the $1s_{\frac{1}{2}}$ shell. 
The full $S_3$ correlation operator therefore increases the calculated radii.
The results obtained for the Carbon and Oxygen isotopes are in the 
following presented as function of the increasing valence neutrons.
Before presenting the results is however worthwhile to remark that
the high order correlation operators generate the interaction of
the valence particles with the closed shell nucleus.
The correlation model treats therefore consistently the ``A'' particles of
the isotopes.
By using generalized linearization approximations and cluster factorization
coefficients we can perform exact calculations. 
In following Tables an over all b=1.76 fm has been used.
  
In Table 1,~3) we give the calculated magnetic moments and rms radii for one-hole and 
for one-particle in $^{16}$O.
 The energy splitting between the ground- and the second (first)
excited states and the electromagnetic transitions for the two isotopes are 
given in Tables 2,~4).

\begin{table}[tbp]\label{b-0}
\hspace{2.5cm}
\begin{tabular}{lcc}
                          & DCM & Exp.~\cite{tun06}\\ \hline
  Magnetic Moment (mm)    & .70 & .7189 \\ \hline
                          & DCM & Exp.~\cite{dej74}\\ \hline
  rms (fm)                & 2.74& 2.73(3)   
\end{tabular}
\caption{Magnetic moment (nm) and rms (fm) of the ground state of $^{15}$O with 
$J=\frac{1}{2}^-;T=\frac{1}{2}$}
\end{table}

\begin{table}[tbp]\label{b-0I}
\hspace{2.5cm}
\begin{tabular}{lcc}
Energy (MeV) & DCM &  Exp.~\cite{tun06} \\ \hline
$ \Delta E_{\frac{1}{2}^-\frac{5}{2}^+}$ & 5.41& 5.24  \\ \hline \hline
 Ratio & DCM & Exp.~\cite{tun06} \\ \hline 
$\frac{BE(E3;\frac{5}{2}^+\to\frac{1}{2}^-)}{BE(M2;\frac{5}{2}^+\to\frac{1}{2}^-)}$
& .15   & .10    \\ \hline
\end{tabular}
\caption{Energy splitting between the ground- and the first
excited state and the corresponding electromagnetic transitions for $^{15}$O.} 
\end{table}

\begin{table}[tbp]
\hspace{2.5cm}
\begin{tabular}{lcc}
                          & DCM & Exp.~\cite{tun06}\\ \hline
  Magnetic Moment (nm)    & -1.88 & -1.89 \\ \hline
                          & DCM   & Exp.~\cite{dej74}\\ \hline
  rms (fm)                & 2.73  & 2.72(3)   
\end{tabular}
\caption{Magnetic moment (nm) and rms (fm) of the ground state of $^{17}$O with 
$J=\frac{5}{2}^+;T=\frac{1}{2}$}
\end{table}

\begin{table}[tbp]\label{b-I}
\hspace{2.5cm}
\begin{tabular}{lcc}
Energy (MeV) & DCM & Exp.~\cite{tun06} \\ \hline 
 $\Delta E_{\frac{1}{2}^+\frac{5}{2}^+}$  & 0.87 & 0.89  \\ \hline \hline
  Transition($e^2fm^4$)  & DCM & Exp.~\cite{tun06} \\ \hline
 $BE(E2;\frac{1}{2}^+\to\frac{5}{2}^+)$ & 2.10   & 2.18$\pm$0.16    \\
\hline 
\end{tabular}
\caption{Energy splitting between the ground- and the first
excited states and the $E_2$ transition for $^{17}$O.} 
\end{table}

In Table~5) we give the calculated results for the energy splitting between 
the ground- and the $2^+$ excited state and the 
corresponding electromagnetic transition for the $^{14}$C with T=1. 
The commutator equations solved are given in Sec.~4.1).
\begin{table}[tbp]\label{b-II}
\hspace{2.5cm}
\begin{tabular}{lccc}
Energy (MeV) & Ref.~\cite{fuj06} & BDCM & Exp.~\cite{ram87} \\ \hline
 $\Delta E_{0^+2^+}$             &  &8.38 & 8.32  \\ \hline \hline
 Transition($e^2fm^4$) & Ref.~\cite{fuj06} & BDCM & Exp.~\cite{ram87} \\ \hline
 $BE(E2;2^+\to0^+)$  & 3.38   & 3.65     &  $3.74\pm.50$ \\
\hline
\end{tabular}
\caption{Calculated energy splitting and $BE(E2;2^+\to0^+)$ transition for $^{14}$C.} 
\end{table}

In Tables 6,~7) we give the results for the energy splitting between the ground- and the 
excited states and the corresponding electromagnetic transitions
 for the $^{15}$C and $^{19}$O with T=$\frac{3}{2}$. The commutator equations used 
are given in Sec.~4.2).
The resulting CMWFs are therefore including (3p) interacting with (4p1h). 
In Tables 8,~9) we give the results for the energy splitting between the ground- and the
excited states and the corresponding $E_2$ transitions
for the $^{16}$C and $^{20}$O with T=2 .Calculations are performed
 by extending the commutator equations given 
in Sec.~4.2) to four valence neutrons.
The resulting CMWFs are then including (4p) interacting with (5p1h). 
Good results have been overall obtained with a neutron effective charge varying between
0.1- to 0.12-$e_n$.  

\section{Conclusion and Outlook}
In this contribution we have investigated the effect of the microscopic 
correlation operators on the exotic structure of the Carbon and
Oxygen isotopes. The microscopic correlation has been separated in short- 
and long-range correlations according to the definition of Shakin.
The short-range correlation has been used to define the effective Hamiltonian
of the model while the long-range correlation is used to calculate the structures and the 
distributions of exotic nuclei. As given in the work of Shakin, only the 
two-body short-range correlation need to be considered in order to derive 
the effective Hamiltonian especially if the correlation is of very short 
range. For the long range correlation operator the three body 
component is important and should not be neglected. Within the three body 
correlation operator, one introduces in the theory a three body interaction 
which compensates for the use of the genuine three body interaction
of the no-core shell model. Within the $S_2$ effective Hamiltonian, good results have been
obtained for the ground state energies and the distributions of
$^3$H, $^3$He, and $^4$He.
The higher order correlation operators $S=3 \cdots n$ have been used 
to calculate the structure and the electromagnetic transitions
of ground and first excited states for the isotopes of Carbon and Oxygen.
By using generalized linearization approximations and cluster factorization
coefficients we can perform expedite and exact calculations.  
Detailed calculations for the  5- and 6-neutron systems are presently under
 investigation and will be shortly reported.   

\begin{table}[tbp]\label{b-III}
\hspace{2.5cm}
\begin{tabular}{lcc}
 Energy(MeV) & DCM  &Exp.~\cite{fuj05} \\ \hline
 $\Delta E_{\frac{1}{2}^-\frac{1}{2}^+}$            &3.15 & 3.10  \\ \hline \hline
 Transition ($e^2fm^2$) & DCM &  Exp.~\cite{fuj05}    \\ \hline
 $BE(E1;\frac{1}{2}^-\to\frac{1}{2}^+)$  & .014   & .018      \\
\hline
\end{tabular}
\caption{Calculated energy splitting and $BE(E1;\frac{1}{2}^-\to\frac{1}{2}^+)$ transition for $^{15}$C.} 
\end{table}

\begin{table}[tbp]\label{b-IV}
\hspace{2.5cm}
\begin{tabular}{lccc}
 Energy (MeV)  & Kuo~\cite{kuo80} & DCM & Exp.~\cite{ajz78} \\ \hline
 $\Delta E_{\frac{5}{2}^+\frac{1}{2}^+}$ (MeV) & - &1.45 & 1.47  \\ \hline \hline
   Transition (W.U. )& Kuo~\cite{kuo80}  & DCM &  Exp.~\cite{ajz78} \\ \hline
 $BE(E2;\frac{1}{2}^+\to\frac{5}{2}^+)$  & .39   & .49     &  $.58\pm.12$ \\
\hline
\end{tabular}
\caption{Calculated energy splitting and $BE(E2;\frac{1}{2}^+\to\frac{5}{2}^+)$ transition for $^{19}$O.} 
\end{table}

\appendix
\section{Definition of the Model CMWFs}
\subsection{CMWFs for Two ``dressed'' Particles}
In the BDCM the degree of linearization applied to the commutator equations
defines the CMWFs of the model.
For A=6 the model space is formed by two valence particle states 
and by the full set of the (3p1h) CMWFs. 
These different components are associated to the following linearization 
mechanisms:
a) In the zero order linearization approximation we retain
only two particle states:
\begin{equation}
\Psi^{2p}(j_1j_2J)=[a^{\dagger}_{j_1}a^{\dagger}_{j_2}]^{JM}|0\rangle
\end{equation}
For the two particles we distinguish between :\\
1) effective valence space which is used to diagonalize the EoM,\\
2) complementary high excited single particle states which are used to compute
the G matrix.\\
b) In the first order linearization approximation we include
in the dynamic theory also the (3p1h) terms.
 These are  generated by the
application of the correlation operator of the third order to the particles
in the open shell states.
Within this linearization approximation the CMWFs of the model are defined by:
\begin{equation}
\Psi^{dressed}(j_1j_2J)=([a^{\dagger}_{j_1}a^{\dagger}_{j_2}]+
 [a^{\dagger}_{j_1}a^{\dagger}_{j_2}]^{J_{12}}[a^{\dagger}_{j_3}a_{j_4}]^{J_{34}})^{JM}|0\rangle.
\end{equation}
The (3p1h) CMWFs are then expanded according to the CFT theory. The expansion
allows to orthonormalize the CMWFs in an easy way. \newline
c) The (4p2h) states which characterize the second order linearization step
are not included in the model space but, linearized, generate
the eigenvalue equation of the model (2p)+(3p1h) states. 
\subsection{CMWFs for n ``dressed' Particles}
The CMWFs for n dressed particles are characterized by the coupling 
 of the n valence particles with the (np(mp-mh)) core excited states.
For both components we introduce cluster transformation coefficients (CFC)
obtained within the CFT theory.
With the use of these coefficients the complex base can be
easily orthogonalized and the CM can be eliminated.
The numerical formulation of these coefficients will be published in short time.

\section{Center-of-mass Correction} 
Before performing the diagonalization of relative Hamilton's operator in the 
CMWFs defined in appendix A) we have to eliminate the spurious center-of-mass states.
We start to compute, following the calculations of Ref.~\cite{bar61},
the percent weights of spurious states in the model wave functions.
These  can be obtained by calculating the energy of the center of mass
according to the following equation:
\begin{equation}\label{eq.b1}
\begin{array}{l}
\displaystyle{
E_R= \int dR\Psi^{\dagger dressed}(j_ij_jJ)(R^2)\Psi^{dressed}(j_ij_jJ)}\\
\displaystyle{
+2\sum_{ij} \int d\vec{r_i}d\vec{r_j}\Psi^{\dagger dressed}(j_ij_jJ)(\vec{r_i}\cdot\vec{r_j})\Psi^{dressed}(j'_ij'_jJ)}.
\end{array}
\end{equation}
In Eq.~(\ref{eq.b1}) 
the calculation of the integrals can be performed by expanding the dressed 
 states in terms of the CFC given in~\cite{tom06} and by considering that 
for two particle states we have:
\begin{equation} \label{eq.b2}
\begin{array}{l}
\langle j_ij_jJ|(\vec{r_i}\cdot\vec{r_j})|j_ij_jJ\rangle\\
=\frac{4\pi}{3}[\hat{j_i} \hat{j_j}]
\left ( \begin{array} {ccc}
 j_i & 1 &j_j\\
-\frac{1}{2}& 0 &\frac{1}{2} \end{array} \right )^2
\left \{
 \begin{array} {ccc}
j_i & j_j &J\\
j_i & j_j &1 \end{array} \right \}
\langle l_i|r|l_j\rangle^2,
\end{array}
\end{equation}
where:
\begin{equation}
\hat{j}=(2j+1).
\end{equation}
By diagonalizing the above operator in the model space we obtain the energy 
of the center of mass.
The overlap with the model space give
 the degree of ``spuriosity'' of the different components.

\begin{table}[tbp]\label{b-V}
\hspace{2.5cm}
\begin{tabular}{lccc}
Energy (MeV) & Ref.~\cite{fuj06} & BDCM & Exp.~\cite{ism04} \\ \hline
  $\Delta E_{0^+2^+}$       & 1.65  & 1.80 & 1.77  \\ \hline \hline
 Transition ($e^2fm^4$) &Ref.~\cite{fuj06}  & BDCM &  Exp.~\cite{ism04}  \\ \hline
 $BE(E2;2^+\to0^+)$  & .85   & .65     &  $.63^{\pm.11(stat)}_{\pm.16(syst)}$ \\
\hline
\end{tabular}
\caption{Calculated energy splitting and $BE(E2;2^+\to0^+)$ transition for $^{16}$C.} 
\end{table}

\begin{table}[tbp]\label{b-VI}
\hspace{2.5cm}
\begin{tabular}{lcc}
 Energy (MeV) & BDCM & Exp.~\cite{jew01}  \\ \hline
 $\Delta E_{0^+2^+}$   &1.45  & 1.47  \\ \hline \hline
 Transition ($e^2fm^4$) & BDCM  & Exp.~\cite{jew01}  \\ \hline
 $BE(E2;0^+\to2^+)$ &    29.3   & 28.   \\
\hline
\end{tabular}
\caption{Calculated energy splitting and $BE(E2;0^+\to2^+)$ transition for $^{20}$O.} 
\end{table}

\section{The Correlated Operators}
In this appendix we give the expectation values of 
operators calculated with dressed (correlated) particles.
In the following we calculate the correlated distributions of
even nuclei and the magnetic moment operator of odd nuclei. 
The distribution of two valence particles is evaluated from
the model CMWFs:
\begin{equation}\label{c1}
\displaystyle{
\langle\tilde{\Psi}_{12}|\rho(r)| \tilde{\Psi}_{12}\rangle =
\sum_{ij}\chi_{ij} \langle \Psi_{ij}|\rho(r)| \Psi_{ij} \rangle +
\sum_{ijkl} \chi_{ijkl}\langle\Psi_{ij}\Psi_{kl}|\rho(r)|\Psi_{ij}\Psi_{kl}\rangle}
\end{equation}

For the magnetic moment operator we have also to calculate matrix
elements between correlated CMWFs. Here for
odd nuclei with one valence particle we calculate by using 
 the corresponding correlated wave functions:
\begin{equation}\label{c2}
\langle\tilde{\Psi}^J|\mu|\tilde{\Psi}^J\rangle
=\chi_{j_1} \langle \Phi_{j_1}|\mu|\Phi_{j_1}\rangle 
+\sum_{ljk}\chi_{ijk}\langle \Phi_{j_l}\Psi_{jk}|\mu|\Phi_{j_l}\Psi_{jk}\rangle
\end{equation}

\end{document}